\RequirePackage{fix-cm}
\documentclass[smallcondensed]{svjour3}     
\smartqed  
\usepackage{graphicx}
\usepackage{csquotes}
\usepackage{hyperref}
\usepackage{cite}
\usepackage{amsmath,amssymb,amsfonts}
\usepackage{algorithmic}
\usepackage{textcomp}
\usepackage{framed,multirow}
\usepackage{float}
\DeclareMathOperator*{\argmin}{arg\,min}
\journalname{Applied Intelligence}
\begin{document}

\title{Automated diagnosis of COVID-19 with limited posteroanterior chest X-ray images using fine-tuned deep neural networks}

\titlerunning{Automated diagnosis of COVID-19...}        

\author{Narinder Singh Punn         \and
        Sonali Agarwal 
}


\institute{Narinder Singh Punn \at
              IIIT Allahabad, Prayagraj, India, 211015 \\
              Tel.: +91-7018466740\\
              \email{pse2017002@iiita.ac.in}           
           \and
           Sonali Agarwal \at
              IIIT Allahabad, Prayagraj, India, 211015 \\
              Tel.: +91-9415647042\\
              \email{sonali@iiita.ac.in}
}


\maketitle

\begin{abstract}
The novel coronavirus 2019 (COVID-19) is a respiratory syndrome that resembles pneumonia. The current diagnostic procedure of COVID-19 follows reverse-transcriptase polymerase chain reaction (RT-PCR) based approach which however is less sensitive to identify the virus at the initial stage. Hence, a more robust and alternate diagnosis technique is desirable. Recently, with the release of publicly available datasets of corona positive patients comprising of computed tomography (CT) and chest X-ray (CXR) imaging; scientists, researchers and healthcare experts are contributing for faster and automated diagnosis of COVID-19 by identifying pulmonary infections using deep learning approaches to achieve better cure and treatment. These datasets have limited samples concerned with the positive COVID-19 cases, which raise the challenge for unbiased learning. Following from this context, this article presents the random oversampling and weighted class loss function approach for unbiased fine-tuned learning (transfer learning) in various state-of-the-art deep learning approaches such as baseline ResNet, Inception-v3, Inception ResNet-v2, DenseNet169, and NASNetLarge to perform binary classification (as normal and COVID-19 cases) and also multi-class classification (as COVID-19, pneumonia, and normal case) of posteroanterior CXR images. Accuracy, precision, recall, loss, and area under the curve (AUC) are utilized to evaluate the performance of the models. Considering the experimental results, the performance of each model is scenario dependent; however, NASNetLarge displayed better scores in contrast to other architectures, which is further compared with other recently proposed approaches. This article also added the visual explanation to illustrate the basis of model classification and perception of COVID-19 in CXR images.
\end{abstract}

\section{Introduction}

Coronaviruses are a large family of viruses that can cause severe illness to the human being. The first known severe epidemic is severe acute respiratory syndrome (SARS) occurred in 2003, whereas the second outbreak began in 2012 in Saudi Arabia with the middle east respiratory syndrome (MERS). The novel coronavirus disease 2019 (COVID-19), began as an outbreak from epicentre Wuhan, People’s Republic of China in late December 2019, and till April 15, 2020, it caused 1,996,681 infections and 127,590 deaths worldwide~\cite{WHO1}. The coronavirus (COVID-19) outbreak was declared a public health emergency of international concern by WHO on January 30, 2020~\cite{WHO2}. On March 11, as the number of COVID-19 cases has increased thirteen times apart from China with more than 118,000 cases in 114 countries and over 4,000 deaths, WHO declared this a pandemic~\cite{punn2020covid}.
Globally, many researchers of medicine, clinical and artificial intelligence areas are trying hard to mobilize preventive action plans for COVID-19 with identified research priorities. Since this disease is highly contagious, the most desirable preventive measure is to identify the infected people to control the spread. Unfortunately, there is no well-known treatment available to cure COVID-19, therefore the identified infected person must be kept in isolation to break the transmission chain as this patient may become the source of community transfer. Till now, the testing kit is the only available option for diagnosis of COVID-19. Unavailability of testing kits due to excessive demand all over the world is a severe problem in the mission against this pandemic. Though several healthcare organizations are claiming for successful development of testing kits, there is a huge gap in demand and supply. The healthcare agencies have accelerated the rate of development of low-cost testing kits, but the inability to diagnose at early-stage and due to exponential growth of COVID-19 cases, medical professionals are bound to rely on other diagnostic measures. 

Clinical studies have shown that most COVID-19 patients suffer from lung infection~\cite{guo2020origin}. Although chest CT is a more effective imaging technique for lung-related disease diagnosis; CXR is preferred because it is widely available, faster and cheaper than CT. Since COVID-19 infection attacks the epithelial cells that line our respiratory tract, hence X-rays images can be used to analyse the lungs to diagnose pneumonia, lung inflammation, abscesses, and/or enlarged lymph nodes~\cite{allen1994eosinophilic}. Due to its easy transmission, developing techniques to accurately and easily identify the presence of COVID-19 and distinguish it from other forms of flu and pneumonia is crucial. 

Biomedical image analysis (segmentation and classification) is an admired area of research to make the healthcare system more promising~\cite{fourcade2019deep}. In this area, advancement in computing infrastructure makes it possible to deploy the deep learning techniques for complex medical image analysis tasks. Recent works have shown that the chest X-rays of patients suffering from COVID-19 depicts certain abnormalities in the radiography~\cite{rajinikanth2020harmony}. For medical image analysis; deep learning techniques, specifically, convolutional neural networks (CNN) are very effective and efficient in feature extraction and learning, hence becoming the most popular choice among researchers~\cite{krizhevsky2012imagenet}. CNNs have been successfully deployed in the analysis of video endoscopy~\cite{gomez2019low} and CT images, and also used for the diagnosis of pediatric pneumonia via chest X‐ray images~\cite{choe2019deep,kermany2018identifying}. Chouhan et al.~\cite{chouhan2020novel} proposed a transfer learning based deep network approach pre-trained on ImageNet~\cite{5206848} for pneumonia detection. Wang et al.~\cite{wang2017hospital} proposed a customized VGG16 model for lung regions identification to classify different types of pneumonia. Later, Ronneburger et al.~\cite{ronneberger2015u} demonstrated the effectiveness of image augmentation with CNN in the presence of a small set of images. In the area of biomedical image classification, Rajpurkar et al.~\cite{rajpurkar2018deep} proposed a dense CNN with 121‐layers to detect several pathologies including pneumonia using chest X‐rays. Lakhani et al.~\cite{lakhani2017deep} obtained an area under the curve (AUC) of 0.95 in pneumonia detection using AlexNet and GoogLeNet along with image augmentation.  Recently, several deep‐learning based COVID‐19 detection techniques have been proposed~\cite{wang2020covid,wang2020deep,WHO3}. Linda et al.~\cite{wang2020covid} introduced a deep CNN, named COVID‐Net for the detection of COVID‐19 cases from the chest X‐ray images. Shuai et al.~\cite{wang2020deep} achieved accuracy, specificity and sensitivity of 89.5\%, 88\% and 87\% respectively for COVID‐19 identification using CT images. 

There are many datasets available about chest X-rays for the detection of pneumonia~\cite{WHO4,irvin2019chexpert,cohen2020covid,WHO5}; but in present research work, COVID-19 X-ray chest images~\cite{cohen2020covid} and Radiological Society of North America (RSNA) images~\cite{WHO5} are utilized to generate all possible samples of chest infection and also to make the study comparable with other research works. The number of COVID-19 infected samples present in this dataset is very limited that may lead to biased outcome, hence the objective of this research is to maximize the learning ability in presence of a small set of positive class samples. For early diagnosis of COVID-19, this article presents the effectiveness of random oversampling and weighted class loss function approaches for unbiased fine-tuned learning (transfer learning) in various state-of-the-art deep learning techniques. 

Rest of the manuscript is organized as follows: recent research articles are discussed in section II, and section III briefs the dataset. Section IV contains the proposed methodology followed by the evaluation metrics in section V. Results are discussed in section VI whereas the last section contains concluding remarks.

\section{Related work}
Due to the ample availability of X-ray machines, disease diagnosis using CXR images are widely used by healthcare experts. In case of any suspect of COVID-19; instead of using test kits, an alternate way to detect pneumonia from the CXR images is required, so that further investigation can be narrowed down for COVID-19 identification. Many studies have been performed on similar ground with several CXR datasets for diagnosis of pneumonia and other complications~\cite{WHO4,irvin2019chexpert,cohen2020covid,WHO5}. These studies also advocate the need of an automated system for quick diagnosis because the manual methods of X-ray analysis are time consuming and unable to serve the purpose due to limited availability of X-ray machine operators or radiologists.

Amid the COVID-19 outbreak, many companies at the global level around the world embraced  a flurry of Artificial Intelligence (AI) based solutions to detect COVID-19 on chest X-ray scans. It is evident that deep learning tools are effectively used to screen mild cases, triage new infections, and monitor disease advancements. This way of diagnosis can reduce the growing burden on radiologists, and also supplant standard nucleic acid tests as the primary diagnostic tool for coronavirus infection. It is also reported that a swab test needs isolation for testing procedure whereas chest X-ray based detection can be easily manageable. Kermany et al. ~\cite{kermany2018identifying} proposed CXR image-based deep learning model to detect pneumonia and classify other diseases using different medical datasets with testing accuracy of 92.80\%. In another similar research, Stephen et al.~\cite{stephen2019efficient} illustrated an efficient deep learning approach for pneumonia classification by using four convolutional layers and two dense layers in addition to classical image augmentation and achieved 93.73\% testing accuracy. Later, Saraiva et al.~\cite{saraiva2019classification} experimented convolutional neural networks to classify images of childhood pneumonia by using a deep learning model with seven convolutional layers along with three dense layers while achieving 95.30\% testing accuracy. Liang and Zheng~\cite{liang2019transfer} demonstrated a transfer learning method with a deep residual network for pediatric pneumonia diagnosis with 49 convolutional layers and two dense layers and achieved 96.70\% testing accuracy. In similar research, Wu et al.~\cite{wu2020predict} focused on convolutional deep neural learning networks and random forest to propose a pneumonia prediction using CXR images and achieved 97\% testing accuracy. 

Afterwards, Narin et al.~\cite{narin2020automatic} proposed a deep convolutional neural network based automatic prediction model of COVID-19 with the help of pre-trained transfer models using CXR images. In this research, authors used ResNet50, InceptionV3 and Inception-ResNetV2 pre-trained models to obtain a higher prediction accuracy for a subset of  X-ray dataset. Apostolopoulos et. al.~\cite{apostolopoulos2020covid} in their study, utilised state-of-the-art convolutional neural network architectures for classifying the CXR images. Transfer Learning was adopted to handle various abnormalities present in the dataset. Two datasets from different repositories have been used to study images of three classes: COVID-19, bacterial/viral pneumonia and normal condition. The article establishes the suitability of the deep learning model with the help of accuracy, sensitivity, and specificity parameters.

In another research, generative adversarial networks (GAN) are used  by Khalifa et al.~\cite{khalifa2020detection} to detect pneumonia from CXR images. The authors addressed the overfitting problem and claimed its robustness by generating more images through GAN. The dataset containing 5863 CXR images of two categories: normal and pneumonia, has been used with typical deep learning models such as AlexNet, GoogLeNet, Squeeznet and Resnet18 to detect pneumonia. This research highlights that the Resnet18 outperformed among other deep transfer models in combination with GAN. Further, Sethy et al.~\cite{sethy2020detection} proposed a deep learning based model to identify coronavirus infections using CXR images. Deep features from CXR images have been extracted and support vect
or machine (SVM) classifier is used to measure accuracy, false positive rate, F1 score, Matthew's correlation coefficient (MCC) and kappa. It is found that ResNet50 in combination with SVM is statistically superior when compared to other models. Later, Bukhari et al.~\cite{bukhari2020diagnostic} also used ResNet-50 CNN architectures on 278 CXR images, partitioned under 3 groups as normal, pneumonia and COVID-19. This approach gave promising results and indicated substantial differentiation of pulmonary changes caused by COVID-19 from the other types of pneumonia. 

Recently, in another research work, an improved ResNet-50 CNN architecture named COVIDResNet has been proposed~\cite{farooq2020covid}, where conventional ResNet-50 model is applied with different training techniques including progressive resizing, cyclical learning rate finding and discriminative learning rates to gain fast and accurate training. The experiment is performed through progressively re-sizing of input images to 128x128x3, 224x224x3 and 229x229x3 pixels, and automatic learning rate selection for fine-tuning the network at each stage. This work claimed to be computationally efficient and highly accurate for multi-class classification. 

A new deep anomaly detection model is developed by Zhang et. al.~\cite{zhang2020covid} for fast and more reliable screening. To evaluate the model performance, CXR image data  of COVID-19 cases and other pneumonia has been collected from two different sources. To eliminate the data imbalance problem in the collected samples, authors proposed a CXR based COVID-19 screening model through anomaly detection task~\cite{pang2019deep}.

Following this context, this article proposes to contribute for early diagnosis of COVID-19 using the state-of-the-art deep learning architectures, assisted with transfer learning and class imbalance learning approaches.

\section{Dataset description}

\begin{figure}
	\centering
	\includegraphics[width=\columnwidth]{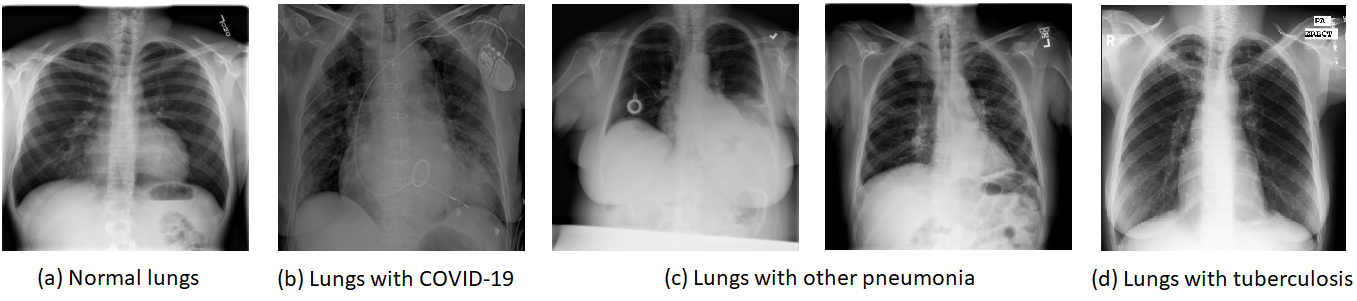}
	\caption{Sample chest radiographs.}
	\label{fig1}
\end{figure}

In this research three datasets are utilized for experiments: COVID-19 image~\cite{cohen2020covid}, Radiological Society of North America (RSNA)~\cite{WHO5} and U.S.  national  library  of  medicine  (USNLM) collected  Montgomery  country - NLM(MC)~\cite{jaeger2014two}. COVID-19 image dataset is a public database of pneumonia cases with CXR images related to COVID-19, MERS, SARS, and ARDS collected by Cohen et al.~\cite{cohen2020covid} from multiple resources available at public domains without infringing patient’s confidentiality (Fig.~\ref{fig1}(b)). It is claimed that this dataset can help to identify characteristics of COVID-19 in contrast to other types of pneumonia; therefore it can play a major role in predicting survival rate. The dataset includes the statistics up to March 25, 2020 consisting of 5 types of pneumonia such as SARSr-CoV-2 or COVID-19, SARSr-CoV-1 or SARS, Streptococcus spp., Pneumocystis spp. and ARDS with following attributes: patient ID, offset, sex, age, finding, survival, view, modality, date, location, filename, doi, url, license, clinical notes, and other notes. 

\begin{table*}[]
	\centering
	\caption{Fused dataset disease class summary details.}
	\label{tab1}
	\begin{tabular}{|l|l|l|l|}
		\hline
		Dataset                   & Findings        & PCXR images & Total samples        \\ \hline
		\multirow{2}{*}{COVID-19} & COVID-19        & 108         & \multirow{2}{*}{153} \\ \cline{2-3}
		& Other pneumonia & 45          &                      \\ \hline
		\multirow{2}{*}{RSNA}     & Normal          & 453         & \multirow{2}{*}{923} \\ \cline{2-3}
		& Other pneumonia & 470         &                      \\ \hline
		\multirow{2}{*}{NLM(MC)}  & Tuberculosis    & 58          & \multirow{2}{*}{138} \\ \cline{2-3}
		& Normal          & 80          &                      \\ \hline
	\end{tabular}
	
\end{table*}

Another dataset utilized in this study is published under RSNA pneumonia detection challenge is a subset of 30,000 examinations taken from the NIH CXR14 dataset~\cite{WHO5}. Out of 30,000 selected images, 15,000 examinations had positive cases of pneumonia and from the remaining 15000 cases, 7500 cases had no findings and other 7500 cases had symptoms other than pneumonia. All these images are annotated by a group of experts including radiologists in two stages. A sample image is shown in Fig.~\ref{fig1}(c). This dataset has been published in two stages. In Stage one, 25,684 training images were considered to test 1,000 images. Later in stage two 1000 testing samples were added to the training set to form the dataset of 26,684 training images and a new set of 3,000 radiographs were introduced for the test. For robust testing and comprehensive coverage of the comparative analysis, NLM(MC)~\cite{jaeger2014two} dataset is also utilized that consists of 138 chest posterior-anterior x-rays samples of tuberculosis and normal cases. A sample image is represented in Fig. 1(d). Table~\ref{tab1} presents the class summary details of the fused dataset resulting from the above discussed datasets which is utilized for training, testing and validation of the proposed approach. The fused dataset is composed of 1214 posteroanterior chest x-ray samples with classes labeled as COVID-19 (108), other pneumonia (515), tuberculosis (58) and normal (533). The generated fused dataset is publicly available~\cite{CXR1}.

\section{Proposed contribution}

The era of artificial intelligence has brought significant improvements in the living society~\cite{makridakis2017forthcoming}. The recent advancements in deep learning have extended its domain in various applications such as healthcare, pixel restoration, visual recognition, signal processing and a lot more~\cite{liu2017survey}. In healthcare domain, the deep learning based image processing approaches for classification and segmentation are applied for faster, efficient, and early diagnosis of the deadly diseases e.g. breast cancer, brain tumor, etc. by using different imaging modalities such as X-ray, CT, MRI,~\cite{shen2017deep} and fused modalities~\cite{punn2020inception} along with its future possibilities. The success of these approaches is dependent on the large amount of data availability, which however is not in the case of automated COVID-19 detection. 

\begin{figure}
	\centering
	\includegraphics[width=\linewidth]{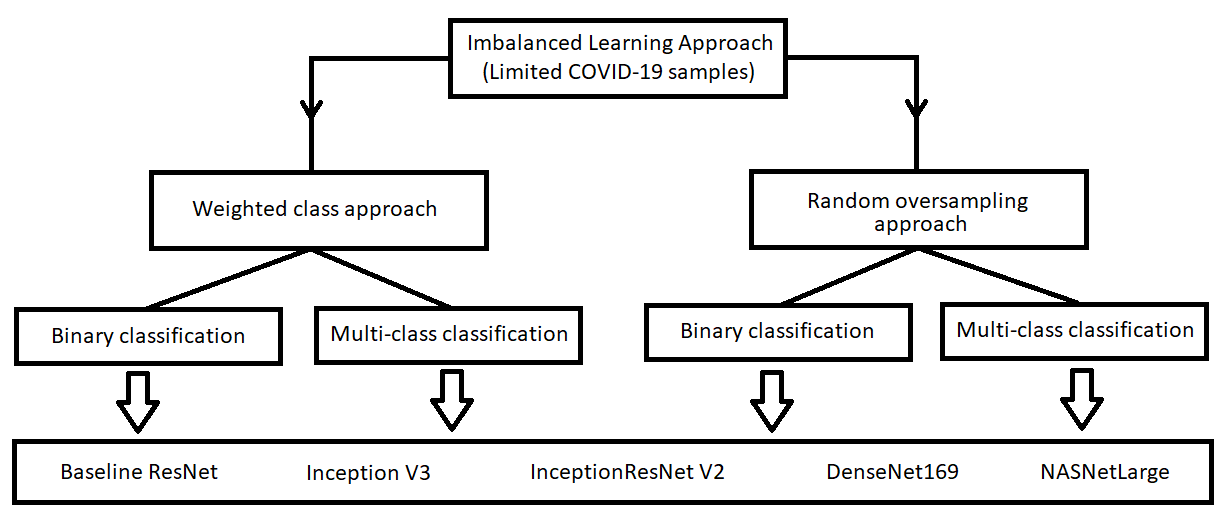}
	\caption{Schematic representation of the proposed components for COVID-19 identification.}
	\label{fig2}
\end{figure}
The main contribution of the work is divided into the components as shown in Fig.~\ref{fig2}. It has two concrete components: data preprocessing and classification. COVID-19 image, RSNA and NLM(MC) datasets are used to generate the final working set. The newly generated dataset contains CXR images of the following classes: coronavirus caused diseases, pneumonia, other diseases and normal cases. Further, binary classification (COVID-19 vs others) and multi-class classification (COVID-19, other types of pneumonia, tuberculosis and normal) are achieved using random oversampling and weighted class loss function approaches for unbiased fine-tuned learning (transfer learning) in various state-of-the-art deep learning approaches such as baseline ResNet, Inception-v3, Inception ResNet-v2, DenseNet169, and NASNetLarge~\cite{szegedy2016rethinking,szegedy2017inception,huang2017densely,zoph2018learning}. The trained models are utilized for identification and classification of COVID-19 in novel samples. Later, visualization techniques are utilized to understand and elaborate the basis of the classification results.

\subsection{Imbalanced Learning Approach}

Class balancing techniques are necessary when minority classes are more important. The dataset used in this research is highly imbalanced which may lead to biased learning of the model. Number of coronavirus infected CXR images are very less compared to other classes, hence class balancing techniques must be insured to smoothen the learning process. This section discusses two approaches to handle the class imbalance problem: weight class approach and random oversampling~\cite{johnson2019survey}.

\subsubsection{Weighted class approach}
In this approach, the intention is to balance the data by altering the weights that each training sample class carries when computing the loss. Normally, each class carries equal weights, but sometimes certain classes with minority samples are required to hold more weights if they are more important because training examples within that class should have a significant effect on the loss function. In the used dataset the coronavirus infected image class samples must be given more weights as they are more significant. In this article, the weights for each class is generated based on the Eq.~\ref{eq1}.

\begin{equation}
{w(c)}=C_{c} . {\frac{\sum\limits_{c=0}^{N}{n_{c}}}{N . n_{c}}}
\label{eq1}
\end{equation}

where $C_c$ is the class constant for a class c, N is the number of classes, and $n_c$ is the number of samples in a class c. The computed class weights are later fused with the objective function (loss function) of the deep learning model in order to heavily penalize the false predictions concerned with the minority samples, which in this case is coronavirus.

\subsubsection{Random oversampling approach}
In this approach, the objective is to increase the number of minority samples by utilizing the existing samples belonging to the minority class. The minority samples are increased until the samples associated with every class become equal. Hence the procedure follows by identifying the difference between the number of samples in majority and minority class. To fill this void of difference, the samples are generated from the randomly selected sample belonging to the minority class by applying certain statistical operations. In this work, the samples of CXR image of COVID-19 positive cases are less as compared to other classes, therefore, these minority class images are randomly oversampled by means of rotation, scaling, and displacement with the objective to achieve equal distribution of classes and accommodate unbiased learning among the deep learning models.

\subsection{Classification Strategy}
Based on the type of data samples availability of CXR images the COVID-19 classification is divided into two following schemes: 
\begin{itemize}
\item Binary Classification - In this classification scheme, the coronavirus positive samples labelled as \enquote{1} (COVID-19) are identified against the rest of the samples labelled as \enquote{0} (non COVID-19 case) which involves other cases e.g. chlamydophila, SARS, streptococcus, tuberculosis, etc., along with the normal cases.
\item Multi-class Classification - In this classification scheme, the aim is to distinguish and identify the COVID-19 samples from the other pneumonia cases along with the presence of tuberculosis and normal case findings. The multi-class classification is performed with three and four classes. The three classes are provided with labels as \enquote{0} being a normal case, \enquote{1} being a COVID-19 case, and \enquote{2} being other pneumonia and tuberculosis cases, whereas four classes are labeled as \enquote{0} being a normal case, \enquote{1} being a COVID-19 case, and \enquote{2} being other pneumonia case and \enquote{3} as tuberculosis case.
\end{itemize}

In both the classification strategies, the deep learning models are trained with the above discussed imbalanced learning approaches using the weighted categorical cross entropy (WCE) loss function as given by Eq.~\ref{eq2.1} and~\ref{eq2}~\cite{koidl2013loss}:

\begin{equation}
f{(s)}_i= \frac{e^{s_i}}{\sum\limits_{C}^{j}{e^{s_i}}}
\label{eq2.1}
\end{equation}
\begin{equation}
WCE=-{\sum\limits_{i}^{C}{w(i).t_i. \log (f{(s)}_i)}}
\label{eq2}
\end{equation}

In categorical cross entropy, the distribution of the predictions (the activations in the output layer, one for each class) is compared with the true distribution only, to ensure the clear representation of the true class as one-hot encoded vector; here, closer the model’s outputs are to that vector, the lower the loss.

\subsection{Data Preprocessing}
In this article, due to the limited samples of posteroanterior chest X-ray images concerned with  positive COVID-19~\cite{cohen2020covid} cases, the data samples are mixed with the other randomly selected CXR images selected from other datasets-, RSNA~\cite{WHO5} and NLM(MC)~\cite{jaeger2014two}. The RSNA and NLM(MC) datasets consists of posteroanterior CXR images covering sample cases labelled as pneumonia and tuberculosis respectively along with normal samples. Table~\ref{tab2} describes the distribution of training, testing, and validation sets using the fused dataset for binary and multi-class classification along with different class imbalance strategies i.e. class weighted loss function that penalizes the model for any false negative prediction and random oversampling~\cite{yap2014application} of minority classes which in this case is COVID-19.

\begin{table*}[]
	\centering
	\caption{Posteroanterior CXR images distribution into training, validation, and test sets from the fused datasets for different problem definitions.}
	\label{tab2}
	\resizebox{\textwidth}{!}{\begin{tabular}{|c|c|c|c|c|c|c|c|c|c|c|c|c|c|c|c|c|c|c|}
			\hline
			\multirow{4}{*}{Findings}                                 & \multicolumn{18}{c|}{Class imbalanced learning strategy}                                                                                                                                                                                                                                                                                                                                                                                                                                                                                       \\ \cline{2-19} 
			& \multicolumn{9}{c|}{Classification with class weighted loss function}                                                                                                                                                                                                 & \multicolumn{9}{c|}{Classification with random oversampling}                                                                                                                                                                                                           \\ \cline{2-19} 
			& \multicolumn{3}{c|}{\begin{tabular}[c]{@{}c@{}}Binary \\ (2 classes)\end{tabular}} & \multicolumn{3}{c|}{\begin{tabular}[c]{@{}c@{}}Multi-class \\ (3 classes)\end{tabular}} & \multicolumn{3}{c|}{\begin{tabular}[c]{@{}c@{}}Multi-class\\ (4 classes)\end{tabular}} & \multicolumn{3}{c|}{\begin{tabular}[c]{@{}c@{}}Binary \\ (2 classes)\end{tabular}} & \multicolumn{3}{c|}{\begin{tabular}[c]{@{}c@{}}Multi-class \\ (3 classes)\end{tabular}} & \multicolumn{3}{c|}{\begin{tabular}[c]{@{}c@{}}Multi-class \\ (4 classes)\end{tabular}} \\ \cline{2-19} 
			& Tr                         & Val                       & Tst                       & Tr                           & Val                         & Tst                        & Tr                          & Val                         & Tst                        & Tr                         & Val                       & Tst                       & Tr                           & Val                         & Tst                        & Tr                           & Val                         & Tst                        \\ \hline
			Normal                                                    & \multirow{3}{*}{906}       & \multirow{3}{*}{90}       & \multirow{3}{*}{110}      & 437                          & 44                          & 52                         & 437                         & 44                          & 52                         & \multirow{3}{*}{960}       & \multirow{3}{*}{90}       & \multirow{3}{*}{110}      & 469                          & 44                          & 52                         & 437                          & 44                          & 52                         \\ \cline{1-1} \cline{5-10} \cline{14-19} 
			Tuberculosis                                              &                            &                           &                           & \multirow{2}{*}{469}         & \multirow{2}{*}{46}         & \multirow{2}{*}{58}        & 47                          & 5                           & 6                          &                            &                           &                           & \multirow{2}{*}{469}         & \multirow{2}{*}{46}         & \multirow{2}{*}{58}        & 437                          & 5                           & 6                          \\ \cline{1-1} \cline{8-10} \cline{17-19} 
			\begin{tabular}[c]{@{}c@{}}Other\\ pneumonia\end{tabular} &                            &                           &                           &                              &                             &                            & 422                         & 41                          & 52                         &                            &                           &                           &                              &                             &                            & 437                          & 41                          & 52                         \\ \hline
			COVID-19                                                  & 88                         & 9                         & 11                        & 88                           & 9                           & 11                         & 88                          & 9                           & 11                         & 960                        & 9                         & 11                        & 469                          & 9                           & 11                         & 437                          & 9                           & 11                         \\ \hline
			Total                                                     & 994                        & 99                        & 121                       & 994                          & 99                          & 121                        & 994                         & 99                          & 121                        & 1920                       & 99                        & 121                       & 1407                         & 99                          & 121                        & 1748                         & 99                          & 121                        \\ \hline
	\end{tabular}}
	
\end{table*}

The CXR images in the aggregated dataset also consists of unwanted artifacts such as bright texts, symbols, varying resolutions and pixel level noise, which necessitates its preprocessing. In order to suppress the highlighted textual and symbolic noise, the images are inpainted with the image mask generated using binary thresholding~\cite{WHO6} as given by equation 4, followed by resizing the images to a fixed size resolution of 331x331x3.

\begin{equation}
  M(x,y)=\begin{cases}
    {max}\_{th}, & i(x,y) \geq {{min}\_{th}} .\\
    0, & \text{otherwise}.
  \end{cases}
 \label{eq3}
\end{equation}

where i(x,y) is an input image, $max_{th}$ and $min_{th}$ are max and min thresholds to design the mask. 
Despite filtering the unwanted information, there is still the possibility of uncertainty at the deep pixel level representation~\cite{hasinoff2014photon}. The denoising or removal of such uncertainty is carried through the adaptive total variation method~\cite{szegedy2015going} while preserving the original distribution of pixel values. 

Let for a given grayscale image $\textit{f}$, on a bounded set $\Omega$ over ${\mathbb{R}}^2$, where $\Omega \subset {\mathbb{R}}^2 $, denoising image u that closely matches to observed image $x=(x_1, x_2)$ $\epsilon$ $\Omega$ - pixels, given as

\begin{equation}
u = \argmin_{u}{\left ( \int_{\Omega}{}(u - f. \ln{u})dx + \int_{\Omega}{}(\omega(x)|\bigtriangledown u|dx) 
 \right)}
\label{eq4}
\end{equation}
where 
$\omega(x) = \frac{1}{1+ k{\mod{G_{\sigma}*\bigtriangledown u}}'}$, $G_{\rho}$ - the Gaussian kernel for smoothing with $\sigma$ variance, k $>$ 0 is contrast parameter and * is convolution operator.

\begin{figure}
	\centering
	\includegraphics[width=\columnwidth] {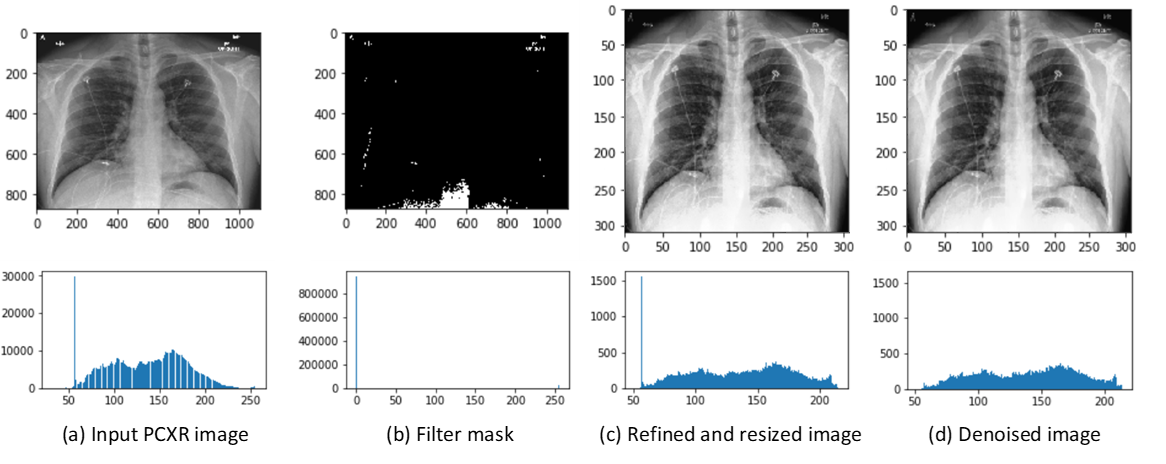}
	\caption{Data preprocessing stages of raw posteroanterior CXR image.}
	\label{fig3}
\end{figure}

Fig.~\ref{fig3} illustrates the data preprocessing stages by considering an instance of COVID-19 case consisting of textual and symbolic artifacts from the generated dataset. The resulting distributed pixels histograms at each stage of preprocessing shown in Fig.~\ref{fig3}, illustrates that the preprocessing approach tends to preserve the original nature of distribution of the pixels while removing the irregular intensities. The preprocessed images are then divided into training, testing, and validation set for training and evaluation of the state-of-the-art deep learning classification models.

\begin{table*}[h!]
	\centering
	\caption{Recent deep learning architectures that are reported with top 5 error rate per year.}
	\label{tab3}
	\begin{tabular}{|c|c|l|c|c|c|}
		\hline
		Architecture                                                  & Year & \multicolumn{1}{c|}{Contribution}                                                                                                                                                & Param. & Dataset                                                                               & E. rate                                                             \\ \hline
		\begin{tabular}[c]{@{}c@{}}Inception-v3\\~\cite{szegedy2016rethinking}\end{tabular} & 2015 & \begin{tabular}[c]{@{}l@{}}Avoided the bottleneck\\    representations,\\ Dimension reduction \\    promotes faster learning.\end{tabular}                                       & 23.6M  & ILSVRC                                                                                & 5.6                                                                 \\ \hline
		\begin{tabular}[c]{@{}c@{}}Inception\\ ResNet-v2\\~\cite{szegedy2017inception}\end{tabular} & 2016 & \begin{tabular}[c]{@{}l@{}}Focused on residual \\    connection rather \\    than filter connection \\    via split and merge\\    approach.\end{tabular}                        & 56M    & ILSVRC                                                                                & 4.9                                                                 \\ \hline
		\begin{tabular}[c]{@{}c@{}}DenseNet169\\~\cite{huang2017densely}\end{tabular}  & 2017 & \begin{tabular}[c]{@{}l@{}}Dense connection blocks\\    for activation flow across \\    the layers.\end{tabular}                                                                & 17.8M  & \begin{tabular}[c]{@{}c@{}}CIFAR-10\\ CIFAR-100\\ CIFAR-10+\\ CIFAR-100+\end{tabular} & \begin{tabular}[c]{@{}c@{}}5.19\\ 19.64\\ 3.46\\ 17.18\end{tabular} \\ \hline
		\begin{tabular}[c]{@{}c@{}}NASNetLarge\\~\cite{zoph2018learning}\end{tabular}                                                   & 2018 & \begin{tabular}[c]{@{}l@{}}Search and transfer the \\    architecture block from \\    small dataset to large\\ dataset,\\ Scheduled drop path\\    regularization.\end{tabular} & 22.6M  & CIFAR-10                                                                              & 2.4                                                                 \\ \hline
	\end{tabular}
\end{table*}

\subsection{Deep learning models}
This section incorporates the state-of-the-art deep learning models utilized in the present research work as shown in  Table~\ref{tab3} along with their respective contribution, parameters, and performance on the standard benchmark datasets. The inception deep convolutional architectures proposed by GoogLeNet are considered as the state-of-the-art deep learning architectures for image analysis and object identification with the basic model as inception-v1~\cite{chen2010adaptive}. Later, this base model was refined by introducing the batch normalization and established as the inception-v2~\cite{szegedy2016rethinking}. In further iterations, additional factorisation was introduced and released as the inception-v3. It is one of the pre-trained models to perform two types of specific tasks: dimensionality reduction using CNN and classification using fully-connected and softmax layers. Since it is originally trained on over a million images consisting of 1,000 classes of ImageNet, its head layers can be retrained for the generated dataset using historical knowledge to reduce the extensive training and computational power. Later, Inception-ResNet-v2 was proposed by Szegedy et al.~\cite{szegedy2017inception}. This hybrid model is a combination of residual connections and a recent version of Inception architecture. It is intended to train very deep convolutional models by the additive merging of signals, both for image recognition and object detection. This network is more robust and learns rich feature representations. Afterwards, DenseNet was proposed by Huang et al.~\cite{huang2017densely}. It works on the concept of reuse, in which each layer receives inputs from all previous layers and yields a condensed model to pass its own feature-maps to all subsequent layers. This makes the network thinner and compact with the fewer number of channels, while improving variation in the input of subsequent layers, and becomes easy to train and highly parameter efficient. Google Brain research team proposed NASNet model based on reinforcement learning search methods~\cite{zoph2018learning}. It creates search space by factoring the network into cells and further dividing it into a number of blocks. Each block is supported by a set of popular operations in CNN models with various kernel size e.g: convolutions, max pooling, average pooling, dilated convolution, depth-wise separable convolutions etc.

\subsection{Fine tuning}
Training deep learning models require (Inception-v3, InceptionResNet-v2, etc.) exhaustive amount of resources and time. While these networks, attain relatively excellent performance on ImageNet~\cite{5206848}, training them on a CPU is an exercise in futility. These CNNs are often trained for a couple of weeks or more using arrays of GPUs to get good results on the complex and complicated datasets. In most deep CNNs the first few convolution layers learn low-level features (edges, curves, blobs) and with progress, through the network, it learns more mid/high-level features or patterns associated with the on-going task. In fine tuning, the aim is to keep or freeze these trained low-level features, and only train the high-level features needed for our new image classification problem.

\begin{figure}[h!]
	\centering
	\includegraphics[width=\linewidth] {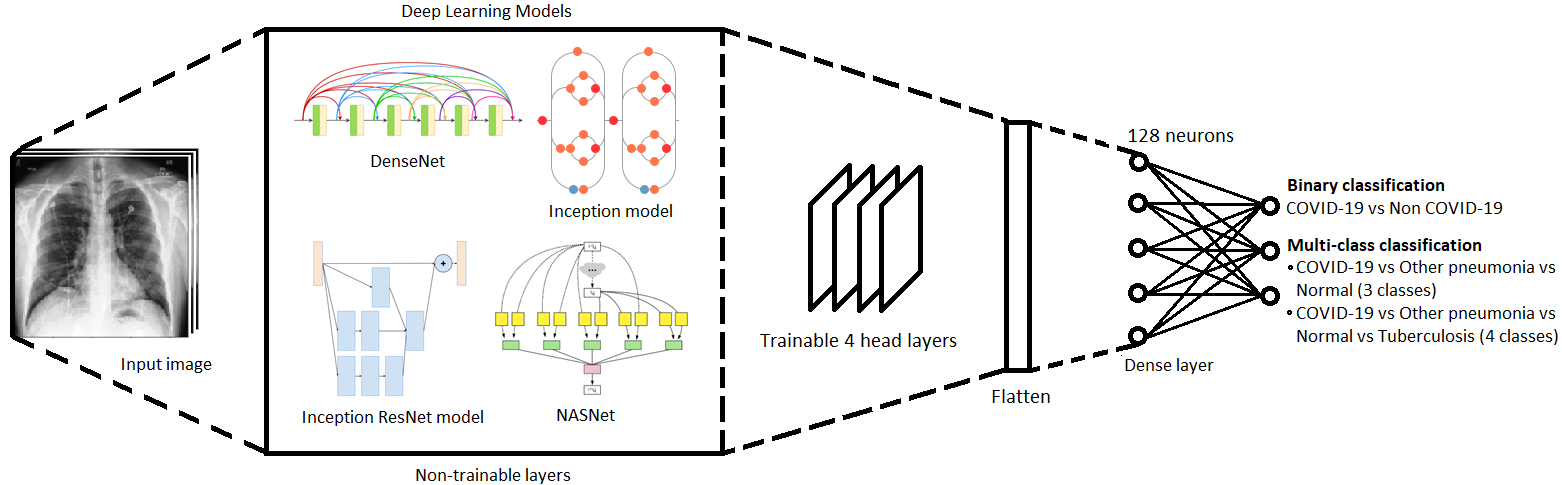}
	\caption{Schematic representation of training framework for deep learning architectures via transfer learning.}
	\label{fig4}
\end{figure}

In this article, the trials of training the deep learning classification models initiates with the baseline residual network that is composed of five residual blocks~\cite{he2016deep}, defined by two convolutions whose rectified linear unit (ReLU) activation is concatenated with the input of the first convolution layer, followed by max-pooling and instance normalization, except the final output layer which uses softmax activation function. Later, transfer learning is utilized on the state-of-the-art architectures discussed in Table~\ref{tab3} to utilize the pre-trained models while fine tuning the head layers to serve the purpose of classifying the COVID-19 samples, as illustrated in Fig.~\ref{fig4}. This follows by enabling the four head layers of the network to adjust its trainable parameters along with the addition of fully connected layers with 128 neurons and 2 or 3 neurons in the output layer depending on binary or multi-class classification, accompanied with the softmax activation function.

\section{Evaluation metrics}
\begin{figure}[h!]
	\centering
	\includegraphics[scale=0.4] {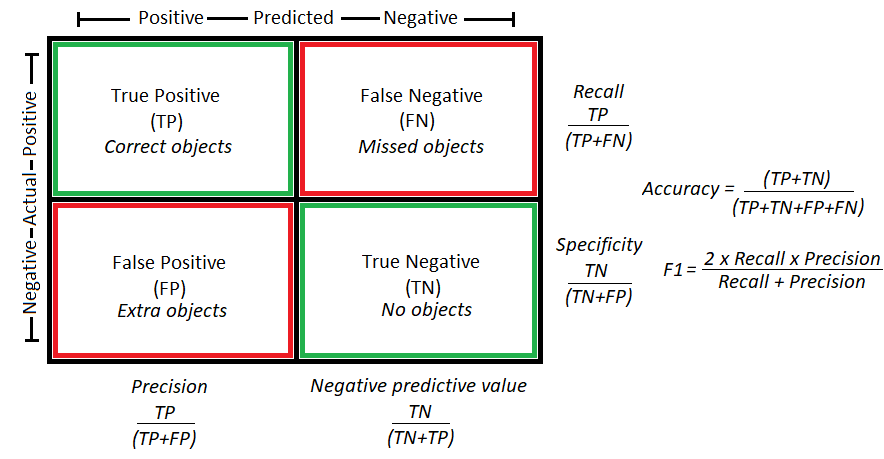}
	\caption{Confusion matrix and performance evaluation metrics.}
	\label{fig5}
\end{figure}

The deep learning models are trained using the training and validation set under consideration of each possible combination of the discussed approaches of class imbalance learning and classification strategy, thereby making four possible scenarios for training a model. Later, the trained models are evaluated on the test set using the standard benchmark performance metrics such as accuracy, precision, recall (selectivity), area under curve (AUC), specificity and F1 score (dice coefficient) as shown in Fig~\ref{fig5}.

\section{Training and testing}
The proposed approach is trained and tested on the above discussed fused dataset using mini-batch gradient descent with learning rate optimizer as adam~\cite{ruder2016overview} via N-Vidia Titan GPU. The training, testing, and validation sets are highlighted in Table~\ref{tab2}. The training phase is assisted with the batch size of 10, 4-fold cross validation, switch normalization~\cite{luo2018differentiable} to adapt to instance or batch or layer normalization, and earlystopping technique that aims to stop the training process if the validation error stops decreasing. For each epoch the above discussed evaluation metrics are computed on the training and validation set to analyse the model's performance and improve the classification results. Later, the test set is utilized to evaluate the results of the proposed approach.

\section{Results and discussion}
\begin{figure}[h!]
	\centering
	\includegraphics[width=\linewidth] {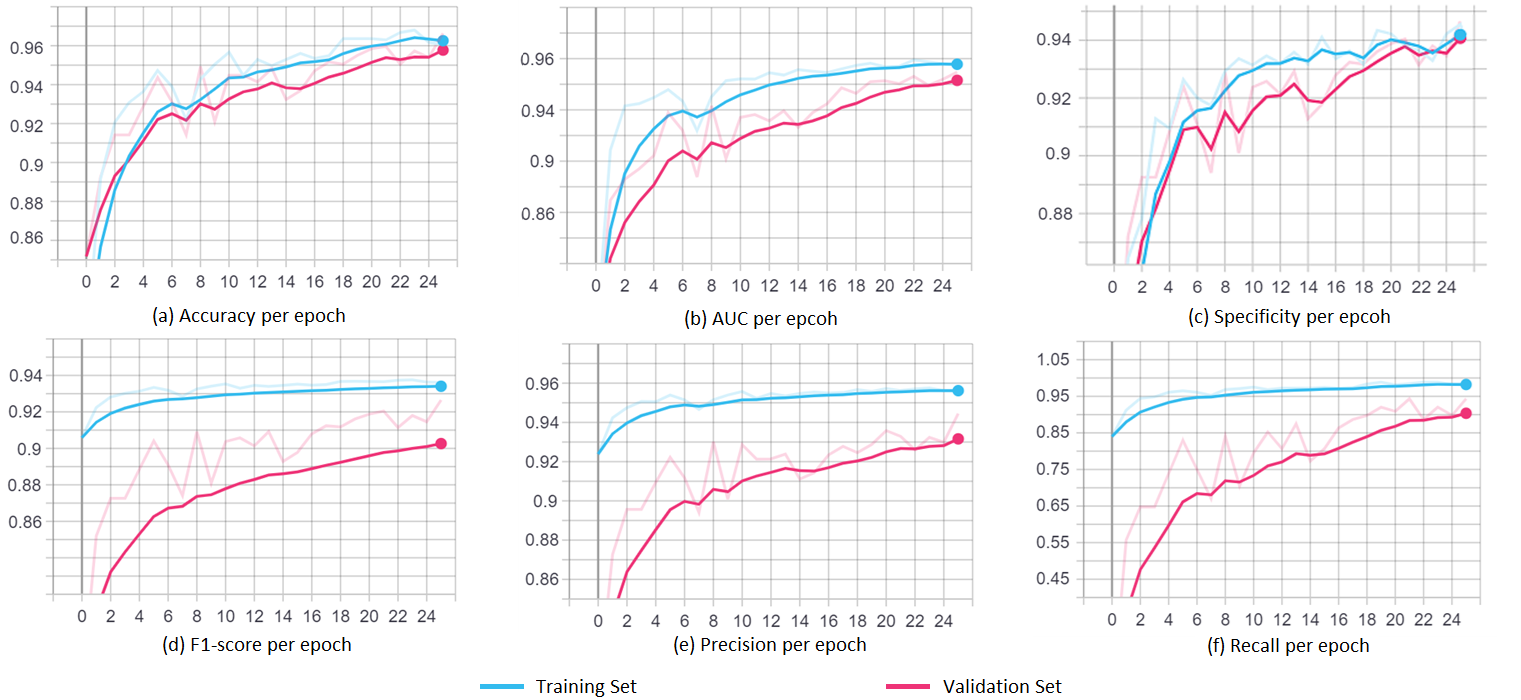}
	\caption{Monitoring every epoch of training with performance curves for the NASNetLarge model with average (a) accuracy, (b) AUC, (c) specificity, (d) F1-score, (e) precision, and (f) recall. The curves are smoothed using moving exponential average.}
	\label{fig6}
\end{figure}

With extensive trials, it is observed that there is no individual model that displayed best performance for all the scenarios in terms of accuracy, precision, recall, specificity, AUC, and F1-score. However, the NASNetLarge mostly claimed the best scores, thus making it best fit for the classification of COVID-19 samples. It is also observed that the results of binary classification (COVID-19 vs non-COVID-19) are better than the multi-class classification (COVID-19 vs other classes). With this it is evident that by grouping other diseases together as non-COVID-19 samples, models can efficiently learn features and patterns belonging to the COVID-19. Fig.~\ref{fig6} shows the average performance curves of the NASNetLarge model evaluated and monitored on the test set for each iteration during the training phase for classification of COVID-19 samples, generated using the tensorboard. Since the models are initialized with the trained weights on the ImageNet and head layers are fine-tuned for classification, the training of the models initiates with descent metrics values. The best scores of the models corresponding to each scenario are highlighted in Table~\ref{tab4} along with the detailed comparative analysis of the results for classifying the COVID-19 samples in terms of discussed evaluation metrics for different scenarios CB, CM$_3$, CM$_4$ RB, RM$_3$ and RM$_4$ where C and R indicates class imbalanced learning approaches, namely weighted class loss functions and random oversampling of the minority class, whereas B indicates the binary classification and M$_3$ (3 classes), M$_4$ (4 classes) indicate multi-class classification schemes. Furthermore, the results obtained are compared with the other recently proposed approaches to detect COVID-19 positive symptoms via X-ray images as shown in Table~\ref{tab5}. It is observed that the proposed approach outperforms other approaches, whereas Apostolopoulos et al.~\cite{apostolopoulos2020covid} achieved similar results with the help of VGG19 model~\cite{simonyan2014very} but with approximately double trainable parameters (144M) as compared to the proposed approach (83M).

\begin{table*}[h!]
\centering
\caption{Comparative analysis of the classification results of the deep learning models for each scenario.}
\label{tab4}
\begin{tabular}{|c|c|c|c|c|c|c|c|}
	\hline
	Model                                                                          & Scenario & Accuracy      & Precision     & Recall        & AUC           & Specifcity    & F1-score      \\ \hline
	\multirow{6}{*}{\begin{tabular}[c]{@{}c@{}}Baseline\\ ResNet\end{tabular}}     & CB       & 0.84          & 0.67          & 0.89          & 0.90          & 0.85          & 0.76          \\ \cline{2-8} 
	& CM$_3$      & 0.77          & 0.62          & 0.89          & 0.90          & 0.77          & 0.73          \\ \cline{2-8} 
	& CM$_4$      & 0.77          & 0.61          & 0.83          & 0.90          & 0.77          & 0.70          \\ \cline{2-8} 
	& RB       & 0.89          & 0.56          & 0.56          & 0.89          & 0.60          & 0.56          \\ \cline{2-8} 
	& RM$_3$      & 0.76          & 0.63          & 0.42          & 0.83          & 0.77          & 0.50          \\ \cline{2-8} 
	& RM$_4$      & 0.76          & 0.65          & 0.43          & 0.82          & 0.73          & 0.52          \\ \hline
	\multirow{6}{*}{\begin{tabular}[c]{@{}c@{}}Inception\\ v2\end{tabular}}        & CB       & 0.88          & 0.77          & 0.79          & 0.93          & 0.89          & 0.78          \\ \cline{2-8} 
	& CM$_3$      & 0.87          & 0.73          & 0.88          & 0.90          & 0.88          & 0.80          \\ \cline{2-8} 
	& CM$_4$      & 0.87          & 0.73          & 0.85          & 0.89          & 0.85          & 0.79          \\ \cline{2-8} 
	& RB       & 0.88          & 0.78          & 0.70          & 0.88          & 0.91          & 0.74          \\ \cline{2-8} 
	& RM$_3$      & 0.87          & 0.90          & 0.81          & 0.91          & 0.90          & 0.85          \\ \cline{2-8} 
	& RM$_4$      & 0.84          & 0.89          & 0.81          & 0.91          & 0.88          & 0.85          \\ \hline
	\multirow{6}{*}{\begin{tabular}[c]{@{}c@{}}Inception\\ ResNet v2\end{tabular}} & CB       & 0.95          & \textbf{0.86} & 0.66          & 0.98          & 0.87          & 0.75          \\ \cline{2-8} 
	& CM$_3$      & 0.92          & 0.85          & 0.92          & \textbf{0.99} & \textbf{0.89} & \textbf{0.88} \\ \cline{2-8} 
	& CM$_4$      & 0.91          & 0.83          & \textbf{0.91} & \textbf{0.98} & \textbf{0.89} & \textbf{0.87} \\ \cline{2-8} 
	& RB       & 0.90          & 0.85          & 0.75          & 0.93          & 0.88          & 0.80          \\ \cline{2-8} 
	& RM$_3$      & 0.92          & \textbf{0.97} & \textbf{0.96} & \textbf{0.99} & 0.94          & \textbf{0.96} \\ \cline{2-8} 
	& RM$_4$      & 0.91          & \textbf{0.95} & \textbf{0.93} & \textbf{0.98} & 0.93          & \textbf{0.94} \\ \hline
	\multirow{6}{*}{\begin{tabular}[c]{@{}c@{}}DenseNet\\ 169\end{tabular}}        & CB       & 0.90          & 0.82          & 0.78          & 0.91          & 0.91          & 0.80          \\ \cline{2-8} 
	& CM$_3$      & 0.87          & 0.87          & \textbf{0.91} & 0.95          & \textbf{0.89} & \textbf{0.89} \\ \cline{2-8} 
	& CM$_4$      & 0.86          & 0.87          & 0.89          & 0.93          & \textbf{0.89} & \textbf{0.88} \\ \cline{2-8} 
	& RB       & 0.95          & \textbf{0.94} & \textbf{0.96} & 0.97          & \textbf{0.95} & 0.95 \\ \cline{2-8} 
	& RM$_3$      & 0.91          & 0.93          & 0.95          & 0.93          & \textbf{0.95} & 0.94          \\ \cline{2-8} 
	& RM$_4$      & 0.92          & 0.94          & 0.95          & 0.94          & 0.95          & 0.94          \\ \hline
	\multirow{6}{*}{\begin{tabular}[c]{@{}c@{}}NASNet\\ Large\end{tabular}}        & CB       & \textbf{0.97} & 0.82          & \textbf{0.91} & \textbf{0.99} & \textbf{0.98} & \textbf{0.86} \\ \cline{2-8} 
	& CM$_3$      & \textbf{0.94} & \textbf{0.89} & \textbf{0.91} & 0.94          & \textbf{0.89} & \textbf{0.90} \\ \cline{2-8} 
	& CM$_4$      & \textbf{0.95} & \textbf{0.88} & 0.89          & 0.92          & \textbf{0.89} & \textbf{0.88} \\ \cline{2-8} 
	& RB       & \textbf{0.98} & 0.87          & 0.90          & \textbf{0.99} & 0.94          & \textbf{0.98}          \\ \cline{2-8} 
	& RM$_3$      & \textbf{0.96} & 0.93          & 0.91          & 0.96          & 0.94          & 0.92          \\ \cline{2-8} 
	& RM$_4$      & \textbf{0.95} & \textbf{0.95} & 0.90          & 0.95          & \textbf{0.94} & 0.92          \\ \hline
\end{tabular}
\end{table*}
\begin{table*}[h!]
	\centering
	\caption{Comparative analysis of the proposed approach with recently proposed strategies.}
	\label{tab5}
	\resizebox{\textwidth}{!}{\begin{tabular}{|p{0.8in}|p{0.5in}|p{0.5in}|l|l|l|l|l|l|}
		\hline
		Authors                                                                                 & Technique                                                               & Network                                                                 & Classes & Accuracy & Precision & Recall                & AUC  & Specificity           \\ \hline
		Yujin et al.~\cite{oh2020deep}                                                                     & \begin{tabular}[c]{@{}l@{}}Fine\\ tuning\end{tabular}                   & \begin{tabular}[c]{@{}l@{}}Res-\\ Net18\end{tabular}                                                               & \multirow{5}{*}{4}       & 0.88     & 0.83      & 0.86                  & -    & 0.96                  \\ \cline{1-3} \cline{5-9}
		Afshar et al.~\cite{afshar2020covid}                                                                    & \begin{tabular}[c]{@{}l@{}}Fine\\ tuning\end{tabular}                   & Capsule                                                                 &        & 0.95     & -         & 0.90                  &      & 0.95                  \\ \cline{1-3} \cline{5-9}
		\textbf{Proposed}                                                                    & \begin{tabular}[c]{@{}l@{}}\textbf{Fine}\\ \textbf{tuning}\end{tabular}                   & \begin{tabular}[c]{@{}l@{}}\textbf{NASNet}\\ \textbf{Large}\end{tabular}                                                                 &        & \textbf{0.95}     & \textbf{0.95}       & \textbf{0.90}                  &  \textbf{0.94}    &  \textbf{0.92}                 \\ \hline
		Wang et al.~\cite{wang2020covid}                                                                     & \begin{tabular}[c]{@{}l@{}}Full\\ training\end{tabular}                 & COVID-Net                                                               & \multirow{5}{*}{3}       & 0.92     & 0.91      & 0.88                  & -    & -                     \\ \cline{1-3} \cline{5-9}
		\begin{tabular}[c]{@{}l@{}}Apostolopoulos\\ et al.~\cite{apostolopoulos2020covid}\end{tabular}                                                                     & \begin{tabular}[c]{@{}l@{}}Fine\\ tuning\end{tabular}                 & VGG19                                                               &        & 0.87     & -      & 0.92                  & -    & 0.98                     \\ \cline{1-3} \cline{5-9}
		\textbf{Proposed}                                                                    & \begin{tabular}[c]{@{}l@{}}\textbf{Fine}\\ \textbf{tuning}\end{tabular}                   & \begin{tabular}[c]{@{}l@{}}\textbf{NASNet}\\ \textbf{Large}\end{tabular}                                                                 &        & \textbf{0.96}     & \textbf{0.93}         & \textbf{0.91}                   & \textbf{0.96}     & \textbf{0.94}                  \\ \hline
		Hall et al.~\cite{hall2020finding}                                                                     & \begin{tabular}[c]{@{}l@{}}Fine\\ tuning\end{tabular}                   & Ensemble                                                               & \multirow{5}{*}{2}       & 0.91     & -      & 0.78                  & 0.94    & 0.93                  \\ \cline{1-3} \cline{5-9}
		\begin{tabular}[c]{@{}l@{}}Apostolopoulos\\ et al.~\cite{apostolopoulos2020covid}\end{tabular}                                                                     & \begin{tabular}[c]{@{}l@{}}Fine\\ tuning\end{tabular}                 & VGG19                                                               &        & 0.98     & -      & 0.92                  & -    & 0.98                     \\ \cline{1-3} \cline{5-9}
		\textbf{Proposed}                                                                    & \begin{tabular}[c]{@{}l@{}}\textbf{Fine}\\ \textbf{tuning}\end{tabular}                   & \begin{tabular}[c]{@{}l@{}}\textbf{NASNet}\\ \textbf{Large}\end{tabular}                                                                 &        & \textbf{0.98}     & \textbf{0.87}         & \textbf{0.90}                   &  \textbf{0.99}    & \textbf{0.98}                  \\ \hline
	\end{tabular}}
\end{table*}

\begin{figure}[]
	\centering
	\includegraphics[width=\linewidth] {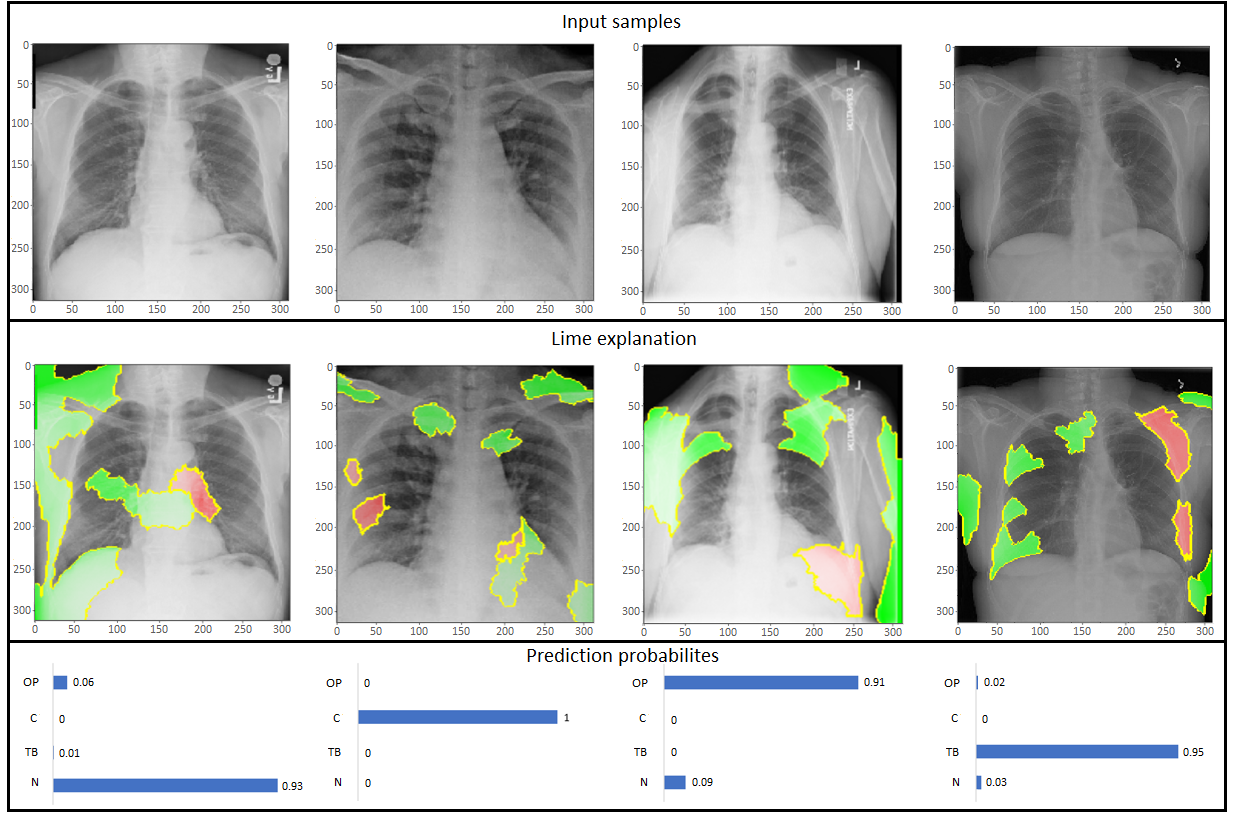}
	\caption{LIME explanation of three distinct class samples using NASNetLarge along with the prediction probabilities of sample being normal (N), COVID-19 (C), and other pneumonia (OP).}
	\label{fig8}
\end{figure}

\section{Visualization}
\begin{figure}[]
	\centering
	\includegraphics[width=\linewidth] {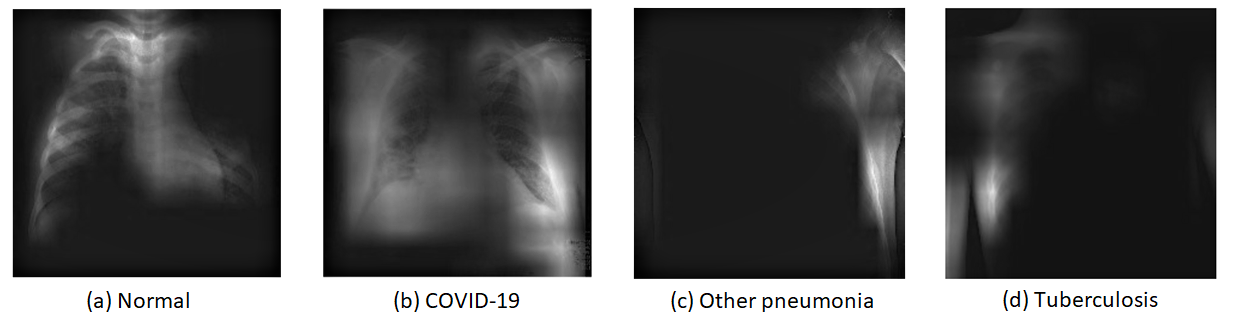}
	\caption{CAM of NASNetLarge model for (a) Normal, (b) COVID-19, and (c) Other pneumonia.}
	\label{fig7}
\end{figure}

Training the model and getting the results is not sufficient unless it is understood that what is triggering the concerned output. To handle this blackbox, visualization techniques assist in illustrating the basis of prediction of the model. There are many visualization techniques for example class activation maps (CAM)~\cite{selvaraju2017grad}, saliency maps (SM)~\cite{simonyan2013deep}, local interpretable model-agnostic explanations (LIME) ~\cite{ribeiro2016should}, and a lot more. In this article, activation maps and LIME techniques are utilized to present the model perception of identifying and classifying the COVID-19 samples from CXR images. CAM aims at understanding the feature space of an input image that influences the prediction, whereas LIME is an innovative explanation technique to represent the model prediction with local fidelity, interpretability and model agnostic. For instance, fine-tuned NASNetLarge architecture is considered to generate LIME explanations for some samples taken from the test set, whereas class activation maps tend to present the patterns learned by the model for classification of samples as shown in Fig.~\ref{fig7}. Fig.~\ref{fig8} presents the LIME technique applied to four samples belonging to four distinct classes as COVID-19, other types of pneumonia, tuberculosis and normal cases. The red and green areas in the LIME generated explanation (Fig.~\ref{fig8}) correspond to the regions that contributed against the predicted class and towards the predicted class respectively.

\section{Conclusion}
This article proposes to leverage the state-of-the-art deep learning models to aid in early identification and diagnosis of COVID-19 virus by using the limited posteroanterior chest X-ray images. Each trained model was evaluated using benchmark performance metrics e.g. accuracy, precision, recall, area under curve, specificity, and F1 score under four different scenarios concerned with imbalanced learning and classification strategy. With extensive trials, it was observed that models achieve different scores in different scenarios, among which NASNetLarge displayed better performance specially in binary classification of COVID-19 samples. The visual representation based on local interpretable model agnostic explanations is utilized to understand the basis of prediction of the model. As an extension to this work more deep learning models and preprocessing techniques can be explored to achieve better results.

\section*{Compliance with Ethical Standards}
\subsection*{Conflict of Interest}
The authors have no conflict of interest to declare.
\subsection*{Ethical approval}
This article does not contain any studies with human participants or animals performed by any of the authors.
\begin{acknowledgements}
We are very much obliged to our institute, Indian Institute of Information Technology Allahabad (IIITA), India and Big Data Analytics (BDA) lab for ensuring the needed resources and support. We also would like to extend our thanks to the colleagues for their valuable guidance and suggestions. The manuscript is also available in the pre-print arXiv with the eprint-2004.11676 under the license Attribution 4.0 International (CC BY 4.0).
\end{acknowledgements}

\bibliographystyle{spmpsci}
\bibliography{reference}

\end{document}